\documentclass{ws-procs975x65}

\begin{document}

\title{Microlensing towards the LMC}

\author{Philippe Jetzer}

\address{Institute of Theoretical Physics, 
University of Z\"urich, 
Winterthurerstrasse 190,
CH-8057 Z\"urich, Switzerland\\
E-mail: jetzer@physik.unizh.ch}

\maketitle

\abstracts{
The nature and the location of the lenses discovered in the
microlensing surveys done so far towards the LMC remain unclear.
Motivated by these questions we computed the optical depth 
for the different intervening populations and
the number of expected events for self-lensing, using a recently drawn 
coherent picture of the geometrical structure and dynamics of the
LMC. By comparing the
theoretical quantities with the values of the observed events it
is possible to put some constraints on the location and the nature
of the MACHOs. 
Clearly, given the large uncertainties and the few
events at disposal it is not yet possible to draw sharp conclusions,
nevertheless we find that up to 
3-4 MACHO events might be due to lenses in LMC, which are most probably
low mass stars, but that hardly all events can be due to
self-lensing. The most
plausible solution is that the events observed so far are due to
lenses belonging to different intervening populations: low mass
stars in the LMC, in the thick disk, in the spheroid and some true
MACHOs in the halo of the Milky Way and the LMC itself.}

\section{Introduction}

The location and the nature of the microlensing events found so
far towards the Large Magellanic Cloud (LMC) is still a matter of
controversy. The MACHO collaboration found 13 to 17 events in 5.7
years of observations, with a mass for the lenses estimated to be
in the range $0.15 - 0.9 ~M_{\odot}$  assuming a
standard spherical Galactic halo\cite{alcock00a} and
derived an optical depth of $\tau= 1.2^{+0.4}_{-0.3} \times
10^{-7}$. The EROS2
collaboration\cite{milsztajn} announced the discovery
of 4 events based on three years of observation but
monitoring about twice as much stars as the MACHO collaboration.
The MACHO collaboration monitored primarily 15 deg$^2$ in the
central part of the LMC, whereas the EROS2 experiment covers a
larger solid angle of 64 deg$^2$ but in less crowded fields. The
EROS2 microlensing rate should thus be less affected by
self-lensing. This might be the reason for the fewer events seen
by EROS2 as compared to the MACHO experiment.

The hypothesis for a self-lensing component
was discussed by several authors 
\cite{sahu,salati,evans98,zhao}.
The analysis of Jetzer et al. \cite{jetzer02} and 
Mancini et al. \cite{mancini} 
has shown that probably the observed
events are distributed among different galactic components (disk,
spheroid, galactic halo, LMC halo and self-lensing). This means that
the lenses do not belong all to the same population and their
astrophysical features can differ deeply one another.

Some of the events found by the MACHO team are most probably due
to self-lensing: the event MACHO-LMC-9 is a double lens with
caustic crossing\cite{alcock00b} and its proper motion is
very low, thus favouring an interpretation as a double lens within
the LMC. The source star for the event MACHO-LMC-14 is double\cite{alcock01b} 
and this has allowed to conclude that the
lens is most probably in the LMC. The expected LMC self-lensing
optical depth due to these two events has been estimated to lie
within the range\cite{alcock01b} $1.1-1.8\times10^{-8}$,
which is still below the expected optical depth for self-lensing
even when considering models giving low values for the optical
depth.
The event LMC-5 is due to a disk lens\cite{alcock01c} and
indeed the lens has even been observed with the HST. 
The other stars which have been
microlensed were also observed but no lens could be detected, thus
implying that the lens cannot be a disk star but has to be either
a true halo object or a faint star or brown dwarf in the LMC
itself.

Thus up to now the question of the location of the observed MACHO events
is unsolved and still subject to discussion. Clearly, with much more
events at disposal one might solve this problem
by looking for instance at their spatial distribution. 
To this end a correct knowledge of the structure
and dynamics of the luminous part of the LMC is essential, and 
we take advantage of the new picture drawn by van der Marel et al. 
\cite{marel01a,marel01b,marel02}.

\section{LMC model}
\label{morphology}

In a series of three 
interesting papers \cite{marel01a,marel01b,marel02}, a
new coherent picture of the geometrical structure and dynamics of
LMC has been given. In the following 
we adopt this model and use 
the same coordinate systems and notations as in van der
Marel. 
We consider an elliptical isothermal flared disk
tipped by an angle $i = 34.7^{\circ} \pm 6.2^{\circ}$ as to the
sky plane, with the closest part in the north-east side. The
center of the disk coincides with the center of the bar 
and its distance from us is $D_{0} = 50.1 \pm
2.5 \, \mathrm{kpc}$. We take a bar mass
$M_{\mathrm{bar}}=1/5\,M_{\mathrm{disk}}$ with
$M_{\mathrm{bar}}+M_{\mathrm{disk}}=M_{\mathrm{vis}}=2.7 \times
10^{9}\, \mathrm{M}_{\odot}$.

The vertical distribution of stars in an isothermal disk is
described by the $\mathrm{sech}^{2}$ function; therefore the
spatial density  of the disk is modeled by:
\begin{equation}
\rho_{\mathrm{d}}=\frac{N\, M_{\mathrm{d}}}{4 \pi\, q\,
R_{\mathrm{d}}^{2}\, \zeta_{\mathrm{d}}(0)}\;
\mathrm{sech}^2\left( \frac{\zeta}{\zeta_{\mathrm{d}}(R)}\right)
 e^{-{\frac{1}{R_{\mathrm{d}}}}\sqrt{
\left(\frac{\xi}{q}\right)^2+
{\eta}^2}}~,
\end{equation}
where $q = 0.688$ is the ellipticity factor,
$R_{\mathrm{d}}=1.54\,\mathrm{kpc}$ is the scale length of the
exponential disk, $R$ is the radial distance from the center on
the disk plane. ${\zeta_{\mathrm{d}}(R)}$ is the \textit{flaring}
scale height, which rises from 0.27 kpc to 1.5 kpc at a distance
of 5.5 kpc from the center\cite{marel02}, and is
given by
$${\zeta_{\mathrm{d}}(R)}=0.27+1.40
\,\tanh \left(\frac{R}{4}\right)~.
$$
$N = 0.2765$ is a normalization factor that takes into
account the flaring scale height.

In a first approach\cite{jetzer02} 
we have described the bar by a Gaussian density profile
following Gyuk et al. \cite{gyuk}, whereas in a following
paper\cite{mancini} we choose, instead, a
bar spatial density  that takes into account its boxy shape\cite{zhao}:

\begin{equation}
\rho_{\mathrm{b}}=\frac{2\,M_{\mathrm{b}}}{\pi^{2} \,
R_{\mathrm{b}}^{2}\,\, \Xi_{\mathrm{b}}}\, e^{
-\left(\frac{\Xi}{\, \Xi_{\mathrm{b}}}\right)^{2}}
\,e^{-\,\frac{1}{R_{\mathrm{b}}^{4}}
\left(\Upsilon^{2}+\,\zeta^{2}\right)^{2}},
\end{equation}
where $\Xi_{\mathrm{b}}=1.2\,\mathrm{kpc}$ is the scale length of
the bar axis,  $R_{\mathrm{b}}=0.44\,\mathrm{kpc}$ is the scale
height along a circular section (for a more detailed discussion
and definition of the coordinate system see \cite{mancini}).

\begin{figure}
 \resizebox{\hsize}{!}{\includegraphics{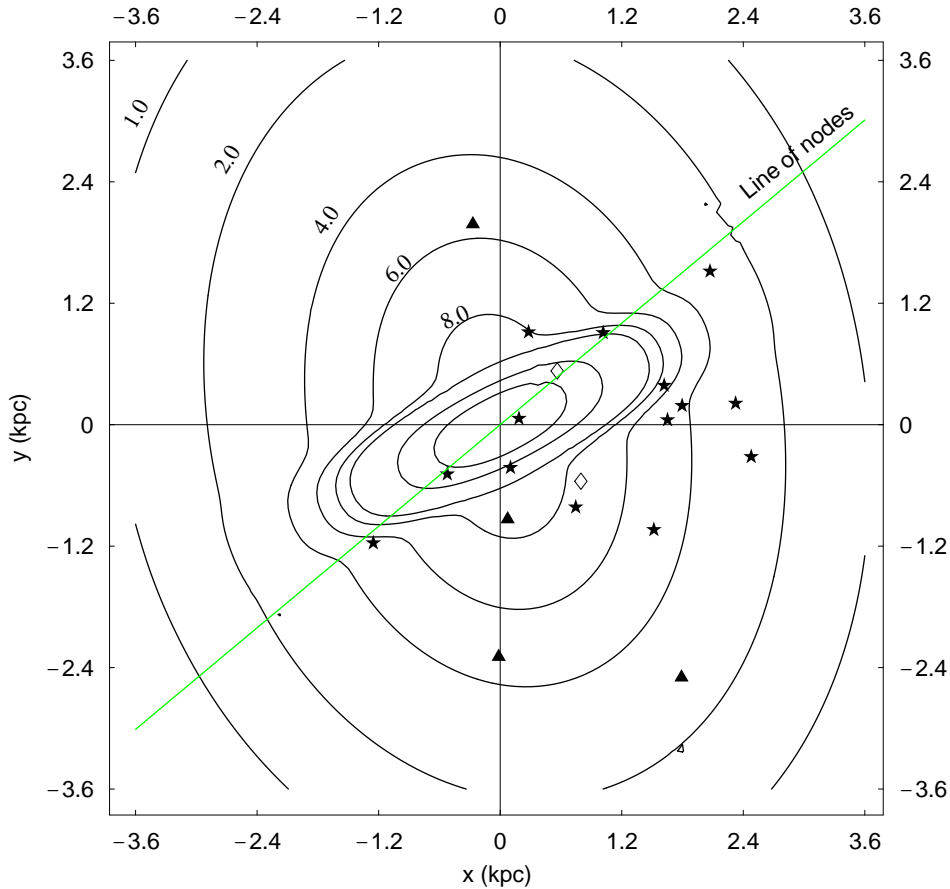}}
 \caption{Projection on the sky plane ($x-y$ plane) of the column density
 of the LMC disk and bar. The numerical values on the contours are in $10^{7}
 \mathrm{M}_{\odot}\mathrm{kpc}^{-2}$ units.
 The three innermost contours correspond to $10$, $20$ and $30\times 10^{7}
 \mathrm{M}_{\odot}\mathrm{kpc}^{-2}$. The locations of the MACHO
 (black stars and empty diamonds) and EROS
 (triangles) microlensing candidates are also shown.}
 \label{column-density-plot}
\end{figure}

The column density, projected on the $x-y$ sky plane is plotted in
Fig. \ref{column-density-plot}, giving a global view of the LMC
shape for a terrestrial observer, together with the positions of
the microlensing events detected by the MACHO (filled 
stars and empty diamonds) and EROS (filled triangles)
collaborations, and the direction of the line of nodes. The
maximum value of the column density, $41.5\times 10^{7}~
\mathrm{M}_{\odot}\,\mathrm{kpc}^{-2}$, is assumed in the center
of LMC. 

We use two different models to describe the halo profile density:
a spherical halo and an ellipsoidal halo. The
values of the parameters have been chosen so that the models have
roughly the same mass within the same radius. In the spherical model
we neglect the tidal effects due to our
Galaxy, and we adopt  a classical pseudo-isothermal spherical  density
profile:
\begin{equation}
\rho_\mathrm{h,S}=\rho_{\mathrm{0,S}}
\left(1+\frac{R^{2}}{a^{2}}\right)^{-1} \theta(R_{\mathrm{t}}-R),
\end{equation}
where $a$ is the LMC halo core radius,  $\rho_{0,\mathrm{S}}$ the
central density, $R_{\mathrm{t}}$ a cutoff radius and $\theta$ the
Heaviside step function. We use $a=2$ kpc.
We fix the value for the mass of the halo within a radius of 8.9
kpc  equal to\cite{marel02} $5.5\times10^{9}~\mathrm{M}_{\odot}$ 
that implies $\rho_{0,\mathrm{S}}$ equal to
$1.76\times10^{7}\,~\mathrm{M}_{\odot}\,\mathrm{kpc}^{-3}$. Assuming
a halo truncation radius\cite{marel02}, $R_{\mathrm{t}}= 15$ kpc, 
the total mass of the halo is $\approx
1.08\times10^{10}\,~\mathrm{M}_{\odot}$.

For the galactic halo we assume a spherical model with density
profile given by:
\begin{equation}
\rho_\mathrm{GH}=\rho_{\mathrm{0}}
\frac{R_{\mathrm{C}}^{2} + R_{\mathrm{S}}^{2}}{R_{\mathrm{C}}^{2}+R^{2}},
\end{equation}
where $R$ is the distance from the galactic center,
$R_{\mathrm{C}} = 5.6 \,\,\mathrm{kpc}$ is
the core radius, $R_{\mathrm{S}} = 8.5 \, \,\mathrm{kpc}$ is the
distance of the Sun from the galactic center and
$\rho_{\mathrm{0}} = 7.9\times 10^{6}\,
~\mathrm{M}_{\odot}\,\mathrm{kpc}^{-3}$ is the mass density in the
solar neighbourhood.

\section{Optical depth} \label{tau}
%
The computation is made by weighting the optical depth with
respect to the  distribution of the source stars along the line of
sight (see Eq.(7) in Jetzer et al. \cite{jetzer02}):
\begin{equation}\label{weightedOD}
\tau = {\frac{4\pi G}{c^{2}}}
{\frac{\int_{0}^{\infty}\left[\int_{0}^{D_{\mathrm{os}}}
{\frac{D_{\mathrm{ol}}(D_{\mathrm{os}}-D_{\mathrm{ol}})}
{D_{\mathrm{os}}}}\rho_{\mathrm{l}}
\,dD_{\mathrm{ol}}\right]\rho_{\mathrm{s}}\,
dD_{\mathrm{os}}}{\int_{0}^{\infty}\rho_{\mathrm{s}}\,
dD_{\mathrm{os}}}}\;.
\end{equation}
$\rho_{\mathrm{l}}$ denotes the mass density of the lenses,
$\rho_{\mathrm{s}}$ the mass density of the sources,
$D_{\mathrm{ol}}$ and $D_{\mathrm{os}}$, respectively, the distance
observer-lens and observer-source.

\begin{figure}
 \resizebox{\hsize}{!}{\includegraphics{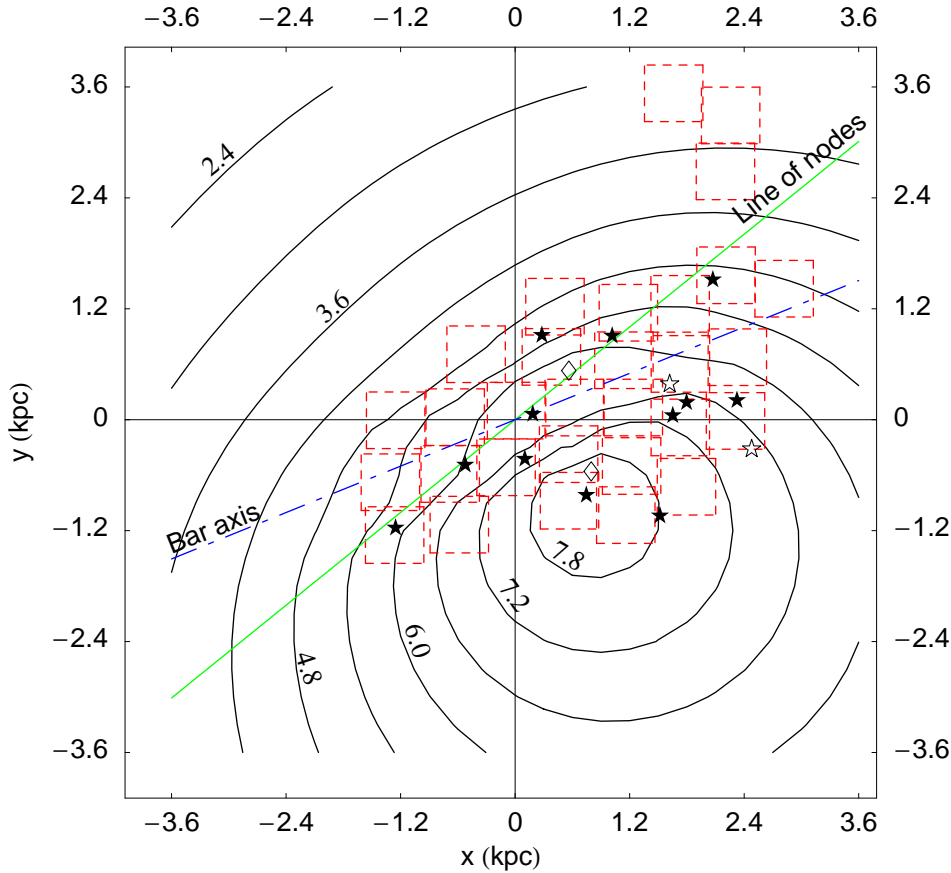}}
 \caption{Spherical halo model: contour map of the optical depth for
 lenses in the LMC halo. The locations of the
 MACHO fields and of the microlensing candidates are also shown.
 The numerical values are in $10^{-8}$ units.}
 \label{LMCHS}
\end{figure}

In Fig. \ref{LMCHS} we report the optical depth
contour maps for lenses belonging to the halo of LMC
in the case of spherical model
in the hypothesis that all the LMC dark halo consists of compact
lenses. The ellipsoidal model leads to similar results\cite{mancini}.
A striking feature of the map is the strong near-far
asymmetry.

For the spherical model, the maximum value of the optical depth,
$\tau_{\mathrm{max,S}} \simeq 8.05 \times 10^{-8}$, is assumed in
a point falling in the field number 13, belonging to the fourth
quadrant, at a distance of $\simeq 1.27$ kpc from the center. The
value in the point symmetrical with respect to the center,
belonging to the second quadrant and falling about at the upward
left corner of the field 82, is $\tau_{\mathrm{S}} \simeq 4.30
\times 10^{-8}$. The increment of the optical depth is of the
order of $\approx 87\%$,  moving from the nearer to the farther
fields. 

\begin{figure}
\resizebox{\hsize}{!}{\includegraphics{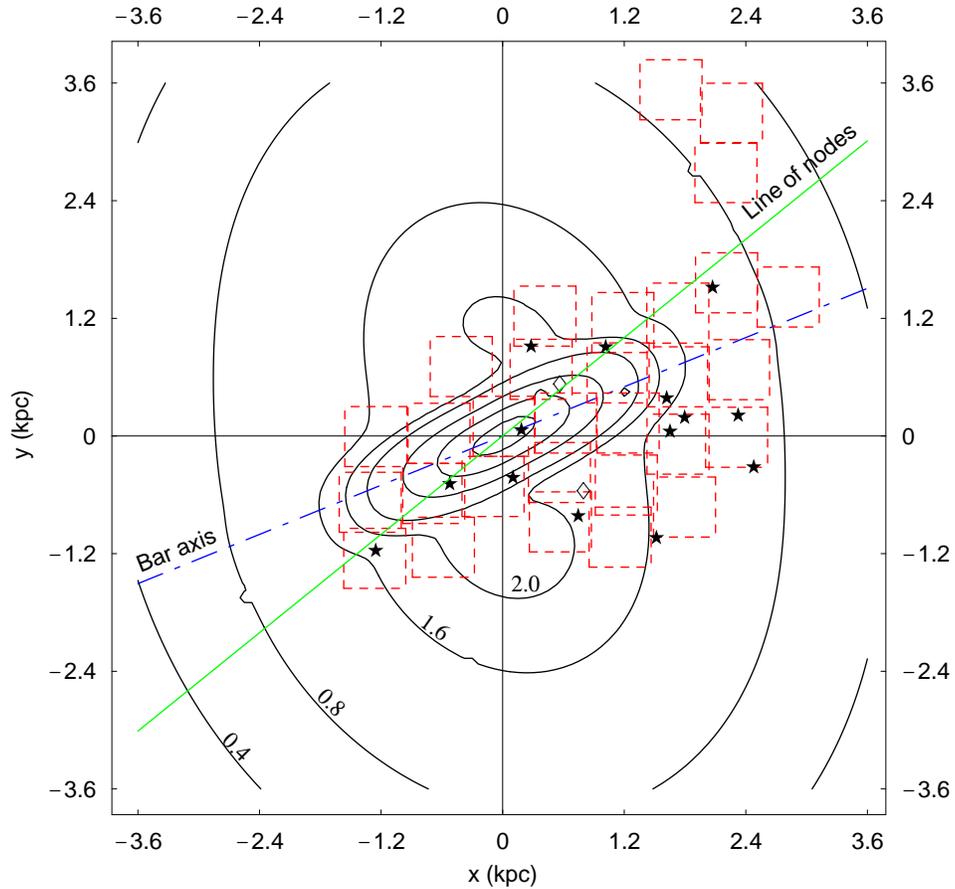}}
\caption{Contour map of the optical depth for self-lensing.  The
 locations of the MACHO fields and of the microlensing candidates
 are also shown. The numerical values are in $10^{-8}$ units.
 The innermost contours correspond to
 values $2.4\times 10^{-8}$, $3.2\times 10^{-8}$, $4.0\times 10^{-8}$
 and $4.6\times 10^{-8}$ respectively.}
\label{SL}
\end{figure}

In Fig. \ref{SL} we report the optical depth contour map for
self-lensing, i. e. for events where both the sources and the
lenses belong to the disk and/or to the bulge of LMC. As expected,
there is almost no near-far asymmetry
and the maximum value of the optical depth, $\tau_{\mathrm{max}}
\simeq 4.80 \times 10^{-8}$, is reached in the center of LMC. The
optical depth then  rapidly decreases, when moving, for instance,
along a line going through the center and perpendicular to the
minor axis of the elliptical disk, that coincides also with the
major axis of the bar. In a range of about only $0.80
\,\mathrm{kpc}$ the optical depth  quickly falls to $\tau \simeq
2 \times 10^{-8}$, and afterwards it decreases slowly to lower
values.
%

\section{Self-lensing event rate}
An important quantity, useful for the physical interpretation of
microlensing events, is the distribution
$\frac{\mathrm{d}\Gamma}{\mathrm{d}\,T_{\mathrm{E}}}$, the differential
rate of microlensing events with respect to the Einstein time
$T_{\mathrm{E}}$. In particular it allows us to estimate the
expected typical duration and their expected number. 
We evaluated the microlensing rate in the self-lensing
configuration, i. e. lenses and sources both in the disk and/or in
the bar of LMC. We have taken into account the transverse motion of
the Sun and of the source stars.
We assumed that, to an observer comoving with the LMC center, the
velocity distribution of the source stars and lenses have a
Maxwellian profile, with spherical symmetry.

In the picture of van der Marel et al. within a distance of
about 3 kpc from the center of LMC, the velocity dispersion
(evaluated for carbon stars) along the line of sight can be
considered  constant, $\sigma_{\mathrm{los}}=20.2\pm .5$ km/s.
Most of the fields of the MACHO collaboration fall within this
radius and, furthermore, self-lensing events are in any case
expected to happen in this inner part of LMC. Therefore, 
we adopted this value, even if we are aware that the
velocity dispersion of different stellar populations in the LMC
varies in a wide range, according to the age of the stellar
population: $\simeq 6$ km/s for the youngest population, until
$\simeq 30$ km/s for the older ones\cite{gyuk}.

We need now to specify the form of the number density. Assuming
that the mass distribution of the lenses is independent of their
position\cite{derujula} 
in LMC ({\it factorization hypothesis}), the lens number
density per unit mass is given by
\begin{equation}
{\frac{\mathrm{d}n_{\mathrm{l}}}{\mathrm{d}
\mu}}={\frac{\rho_{\mathrm{d}}+\rho_{\mathrm{b}}}{\mathrm{M}_{\odot}}}\,
{\frac{\mathrm{d} n_{0}}{\mathrm{d} \mu}},
\end{equation}
where we use ${\frac{\mathrm{d} n_{0}}{\mathrm{d} \mu}}$  as given
in Chabrier\cite{chabrier} ($\mu = M/M_{\odot}$). 
We consider both the power law and the
exponential initial mass functions\footnote{We have used the same
normalization as in Jetzer et al. \cite{jetzer02}
with the mass varying in the range
0.08 to 10 M$_{\odot}$.}. However, we find that our results do not
depend strongly on that choice and hereafter, we will discuss the
results we obtain by using the exponential IMF only.

Let us note that, considering the experimental conditions for the
observations of the MACHO team, we use as  range 
for the lens masses  $0.08 \le\mu \le 1.5$. The lower limit
is fixed by the fact that the lens must be a star in LMC, while
the upper limit is fixed by the requirement that the lenses are
not resolved stars\footnote{We have checked that the results are
insensitive to the precise upper limit value.}.

We compute the ``field exposure'', 
$E_{\mathrm{field}}$, defined, as in Alcock et al. \cite{alcock00a}, 
as the product of the number of distinct light curves per
field by the relevant time span, paying attention to eliminate
the field overlaps; moreover we calculate the distribution
${\frac{\mathrm{d}\Gamma}{\mathrm{d}T_{\mathrm{E}}}}$ along the line
of sight pointing towards the center of each field. In this way we
obtain the number of  expected events for self-lensing, field
by field, given by
\begin{equation}
N_{\mathrm{SL,field}}=E_{\mathrm{field}}\int_{0}^{\infty}\,
{\frac{\mathrm{d}\Gamma}{\mathrm{d}T_{\mathrm{E}}}}\, 
E(T_{\mathrm{E}})\, \mathrm{d}\,T_{\mathrm{E}} \; ,
\end{equation}
where $E(T_{\mathrm{E}})$ is the detection efficiency.

Summing over all fields we find that the expected total number of self-lensing events is 
$\sim 1.2$, while we would get $\sim 1.3$ with the the double power law IMF; in both cases
altogether 1-2 events\cite{mancini}. Clearly, taking also into account the
uncertainties in the parameter used following the van der Marel
model for the LMC the actual number could also be somewhat higher
but hardly more than our upper limit 
estimate of about 3-4 events given in\cite{jetzer02}.

\subsection{Self-lensing events
discrimination}

%
It turns out that, in the framework of the LMC
geometrical structure and dynamics outlined above, a
suitable statistical analysis allows us to exclude from the
self-lensing population a large subset of the detected events. To
this purpose, assuming  all the 14 events as self-lensing, we
study the  scatter plots  correlating the self-lensing expected
values of some meaningful microlensing variables with the measured
Einstein time or with the self-lensing optical depth. In this way
we can show that a large subset of events is clearly incompatible with
the self-lensing hypothesis.

We have calculated the self-lensing distributions
$\left({\frac{\mathrm{d}\Gamma}{\mathrm{d}T_{\mathrm{E}}}}\right)_{\varepsilon}$
of the rate of microlensing events with respect to the Einstein
time $T_{\mathrm{E}}$, along the lines of sight towards the 14
events found by the MACHO collaboration, in the case of a Chabrier
exponential type IMF. 
With these distributions we have calculated the modal
$T_{\mathrm{E,mod}}$, the median $T_{\mathrm{E},\,50\,\%}$ and the
average $<T_{\mathrm{E}}>$ values of the Einstein time.

\begin{figure}
\resizebox{\hsize}{!} {\includegraphics{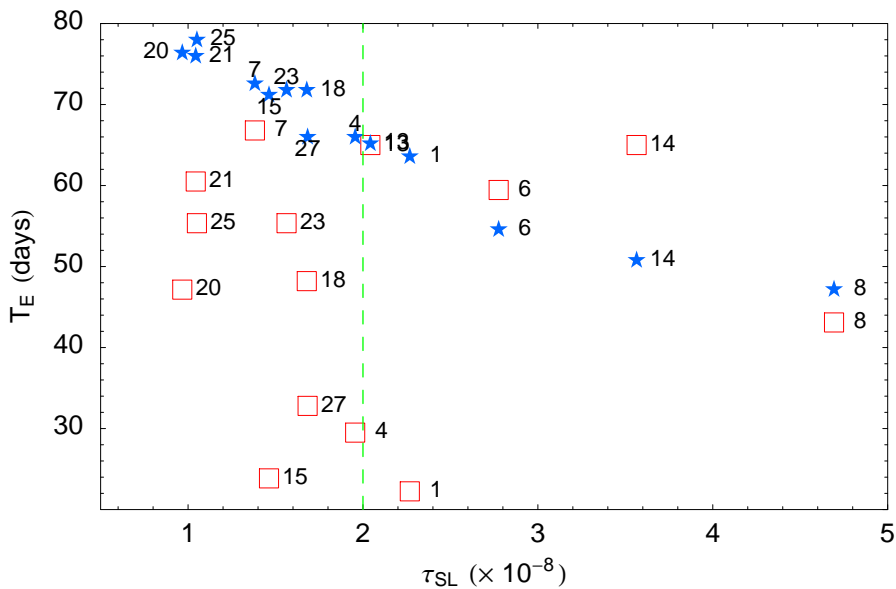}}
{\caption{Scatter plot of the observed (empty boxes) values of the
Einstein time and of the expected values of the median
$T_{\mathrm{E},50\, \%}$ (filled stars), with respect to the
self-lensing optical depth evaluated along the directions of the
events.} \label{tevstau}}
\end{figure}

In Fig. \ref{tevstau} we report on the $y$--axis the observed
values of $T_\mathrm{E}$ (empty boxes) as well as the expected
values for self-lensing of the \emph{median}
$T_{\mathrm{E}\,,50\,\%}$ (filled stars) evaluated
\emph{along the directions of the events}. On the $x$--axis we
report the value of the self-lensing optical depth calculated
towards the event position; the optical depth is  growing going
from the outer regions towards the center of LMC according to the
contour lines shown in Fig. \ref{SL}. An interesting feature
emerging clearly is the \emph{decreasing} trend of the  expected
values of the median $T_{\mathrm{E}\,,50\,\%}$, going from the
outside fields with low values of $\tau_{\mathrm{SL}}$ towards the
central fields with higher values of $\tau_{\mathrm{SL}}$. The
variation of the stellar number density and the flaring of the LMC
disk certainly contributes to explain this behaviour.

We now tentatively identify two subsets of events: the nine
falling outside the contour line $\tau_{\mathrm{SL}} = 2 \times
10^{-8}$ of Fig. \ref{SL} and the five falling inside. In the
framework of van der Marel et al. LMC geometry, this  contour line
includes almost fully the LMC bar and two ear shaped inner regions
of the disk, where we expect self-lensing events to be located with
higher probability.

We note that, at glance, the two clusters have a clear-cut
different collective behaviour: the measured Einstein times of the
first $9$ points fluctuate around a median value of 48 days, very
far from the expected values of the median $T_\mathrm{E}$, ranging
from 66 days to 78 days, with an average value of 72 days. On the
contrary, for the last 5 points, the measured Einstein times
fluctuate around a median value of 59 days, very near to the
average value 56 days of the expected medians, ranging from 47
days to 65 days. Let us note, also, the somewhat peculiar position
of the event LMC--1, with a very low value of the observed
$T_{\mathrm{E}}$; 
most probably this event is homogeneous to the set at left of the
vertical line in Fig. \ref{tevstau} and it has to be included in
that cluster.

This plot gives a first clear evidence that, in the framework of
van der Marel et al. LMC geometry, the self-lensing events have
to be searched among the cluster of events with
$\tau_{\mathrm{SL}}\,>\, 2\times 10^{-8}$, and at the same time
that the cluster of the $9$ events including LMC--1  belongs, very
probably, to a different population.

Moreover, when looking at the spatial distribution of the events
one sees a clear near-far asymmetry in the van der Marel
geometry; they are concentrated along the extension of the bar and
in the south-west side of LMC. Indeed, we have performed a statistical 
analysis of the spatial distribution of the events, which clearly 
shows that the observed asymmetry is greater than the one expected
on the basis of the observational strategy\cite{mancini}.

\section{Conclusions} 

We have presented the results of microlensing survey
towards LMC by using the new picture 
of LMC given by van der Marel et al. \cite{marel01a}.
One interesting feature, that clearly emerges in this framework
by studying the microlensing signature we expect to find,
is an evident near--far asymmetry of the optical depth for 
lenses located in the LMC halo. 
Indeed, similarly to the case of M31 \cite{crotts,jetzer}, 
and as first pointed out by Gould\cite{gould93},
since the LMC disk is inclined, the optical depth is higher along 
lines of sight passing through larger portions of the LMC halo.
Such an asymmetry is not expected, on the contrary,
for a self-lensing population of events. 
What we show is that, indeed, a spatial asymmetry that goes
beyond the one expected from the observational strategy alone,
and that is coherent with that expected because of the inclination
of the LMC disk, is actually present. With the care suggested
by the small number of detected events on which this analysis
is based, this can be looked at, as yet observed by Gould\cite{gould93}, as a
signature of the presence of an extended halo around LMC.

As already remarked, any spatial asymmetry is \emph{not}
expected for a self-lensing population of events,
so that what emerges from this analysis can be considered
as an argument to exclude it. 
 
Furthermore, keeping in mind the observation\cite{evans}
that the timescale distribution of 
the events and their spatial variation across the
LMC disk offers possibilities of identifying the dominant lens
population, we have carefully characterized the ensemble 
of observed events under the hypothesis that
all of them do belong to the self-lensing population.
Through this analysis we have been able
to identify a large subset of events that
can not be accounted as part of this population.
Again, the small amount of events at disposal does not yet
allow us to draw sharp conclusions, although, the various arguments 
mentioned above are all consistent among them and converge quite clearly
in the direction of excluding self-lensing as being the major
cause for the events.

Once more observations will be available, as will hopefully
be the case with the SuperMacho experiment under way \cite{stubbs}, 
the use of the above outlined  methods can bring to a definitive answer to the problem
of the location of the MACHOs and thus also to their nature.

------------------------------------------------------------------

\bibliographystyle{aa}

\end{document}